\documentclass[aps,prd,preprint,floatfix,nofootinbib,longbibliography,a4paper]{revtex4-1}
\pdfoutput=1
\usepackage{color}
\usepackage{graphicx}
\usepackage{notoccite}

\usepackage{textcomp}
\usepackage{subfigure}
\usepackage{feynmp}
\usepackage{amsmath,amssymb,array}
\usepackage{enumerate}
\usepackage{hhline}
\newcommand{\mathsym}[1]{{}}

\def\10{$SO(10)$}

\newcommand{\ba}{\begin{array}}
\newcommand{\ea}{\end{array}}
\newcommand{\be}{\begin{equation}}
\newcommand{\ee}{\end{equation}}
\newcommand{\beqa}{\begin{eqnarray}}
\newcommand{\eeqa}{\end{eqnarray}}
\def\321{$SU(3)\times SU(2)\times U(1)$}
\def\b126{$\overline{126}$}

\usepackage{accents}
\usepackage{tikz}
\usepackage{lmodern}
\usetikzlibrary{decorations.pathmorphing}
\tikzset{snake it/.style={decorate, decoration=snake}}
\usepackage{fancybox}
\expandafter\ifx\csname package@font\endcsname\relax\else
\expandafter\expandafter
\expandafter\usepackage
\expandafter\expandafter
\expandafter{\csname package@font\endcsname}%
\fi
\DeclareFontFamily{OT1}{pzc}{}
\DeclareFontShape{OT1}{pzc}{m}{it}%
             {<-> s * [0.900] pzcmi7t}{}
\DeclareMathAlphabet{\mathscr}{OT1}{pzc}%
                                 {m}{it}

\usepackage{epsfig}
\usepackage{graphicx}
\usepackage{amsmath}
\usepackage{amssymb}
\usepackage{epstopdf}
\usepackage{xcolor}
\usepackage{subfigure}
\usepackage{bm}
\usepackage{dsfont}
\usepackage{hyperref}
\usepackage{braket}
\usepackage{todonotes}
\usepackage{physics}
\usepackage{comment}
\usepackage{tikz}
\usepackage{pgfplots}
\pgfplotsset{compat=newest}
\newcommand{\bea}{\begin{eqnarray}}
\newcommand{\eea}{\end{eqnarray}}

\allowdisplaybreaks
\begin{document}
\title{Stress energy correlator in de Sitter space-time : its conformal masking or growth in connected Friedmann universes}
\bigskip

 \author{Ankit Dhanuka}
 \email{ankitdhanuka@iisermohali.ac.in}
 \affiliation{Department of Physical Sciences, Indian Institute of Science Education \& Research (IISER) Mohali, Sector 81 SAS Nagar, Manauli PO 140306 Punjab India.}

 \author{Kinjalk Lochan}
 \email{kinjalk@iisermohali.ac.in}
 \affiliation{Department of Physical Sciences, Indian Institute of Science Education \& Research (IISER) Mohali, Sector 81 SAS Nagar, Manauli PO 140306 Punjab India.}
 
 \bigskip
\bigskip
\begin{abstract}
Semiclassical Physics in gravitational scenario, in its first approximation (1st order) cares only for the expectation value of stress energy tensor and ignores the inherent quantum fluctuations thereof. In the approach of stochastic gravity, on the other hand, these matter fluctuations are supposed to work as the source of geometry fluctuations and have the potential to render the results from 1st order semiclassical physics irrelevant. We study the object of central significance in stochastic gravity, i.e. the noise kernel,  for a wide class of Friedmann space-times. Through an equivalence of quantum fields on de Sitter space-time and those on generic Friedmann universes, we obtain the  noise kernel through the correlators of Stress Energy Tensor (SET) for fixed co-moving but large physical distances. We show that in many Friedmann universes including  the  expanding universes, the initial quantum fluctuations, the universe is born with, may remain invariant and important even at late times. Further, we  explore the cosmological space-times where even after long times the quantum fluctuations remain strong and become dominant over large physical distances, which the matter driven universe is an example of. The study is carried out in minimal as well as non-minimal interaction settings. Implications of such quantum fluctuations are discussed.
\end{abstract} 

\maketitle

\section*{Introduction}
Study of quantum matter over a classical geometry has given rise to many novel and intriguing features like Hawking radiation, Unruh effect, gravitational particle creation etc. 
\cite{Fulling:1972md},\cite{Hawking:1974sw},\cite{Davies:1974th},\cite{Davies:1976th}. In absence of a complete theory of quantum gravity, semiclassical physics,  which is a 'first' order quantum correction to the classical general relativity,  is the only available methodology capturing the interplay between concepts of quantum mechanics (such as Hilbert space, wavefunction superposition) and those of general relativity (such as general covariance, geodesic distance etc.). Since the inception of the idea of using the quantum expectation value on the right hand side of the Einstein field equations, there has been some level of discomfort regarding its operational status 
\cite{Phillips:2000bp, Hu:2008rga}, more particularly its handling of the inherent quantum fluctuations. One can envisage that in situations where fluctuations tend to grow,
the usage of  quantum  expectation values of the stress tensor alone would not remain justifiable for any physical interpretation. Thus in the scenarios, where significant physical insights depend upon the geometrical structure obtained through the expectation values, it is worthwhile to address  the contributions from fluctuations as well. Gravitational particle creation during early inflationary phase of the universe is such an avenue, where the expectation value for the dynamics-driving field (called inflaton) sets up the expansion of the universe, which in turn creates particles from the perturbation field  \cite{Brandenberger:1984cz,Parker:2009uva, gravitation, Akhmedov:2013vka}. This semiclassical program has been really successful in predicting various novel features of early universe, many of which have also received observational vindication \cite{Akrami:2018odb}. Further, there have been attempts of using different kinds of accelerating phase to obtain or explain various features of the early universe's spectra of quantum predictions \cite{Agullo:2008ka, Brandenberger:2016vhg, Agullo:2016hap}.  Such efforts, are expected to be obtained from different quantum states and subsequent evolution of stress energy expectation values in these states. Further, the late time acceleration, in many discussions \cite{Carroll:2000fy, Burgess:2003jk, Hu:2008rga,Carlip:2018zsk} is also attributed to the quantum character of the stress energy tensor. Therefore, it is imperative to analyse if the semiclassical study,  directing such physical discourse is stable under fluctuations in stress energy tensor.

It is natural to expect that any good quantum gravity theory would yield the known classical results in some appropriate classical limit (in a spirit similar to Ernfest's theorem), apart from capturing the quantum fluctuations. Further, if one seeks an $n^{th}$ order quantum correction to the classical theory, motivated from the semiclassical approach, it is also natural to expect that the quantum gravity theory should return these values in the properly considered limits. Stochastic gravity is one such approach to incorporate quantum fluctuations to the semiclassical gravity, where one considers the effect of quantum fluctuations in matter fields on the classical geometry of space-time. This is accomplished using the Einstein Langevin Equation 
\cite{Phillips:2000bp, Hu:2008rga} in which a central object in the form of {\it Noise Kernel} which works as a stochastic source in addition to the quantum averaged stress tensor. The noise kernel is the vacuum expectation value of symmetrized stress-energy bi-tensor for a quantum field in curved spacetime. As a result,  these fluctuations, if strong enough, will lead to fluctuations in the geometry too \cite{Phillips:2000bp, Hu:2008rga}.  Such induced fluctuations have played an important role in many studies involving backreaction. Backreaction problems in gravity and cosmology, for example, have been addressed \cite{Phillips:2000bp, Hu:2008rga, Satin:2012wj, Satin:2016pjy, Satin:2018zsr}  using Einstein-Langevin equations.


The role of quantum fluctuations themselves in the context of early universe is quite vital. The vacuum fluctuations are understood to form the seed of the modern day galaxy clusters during the inflationary epoch \cite{Brandenberger:1984cz,Parker:2009uva, gravitation, Akhmedov:2013vka}. Present understanding  and observational signatures \cite{Akrami:2018odb} suggest that the universe was born in a very near de Sitter configuration. The de Sitter space-time, like flat spacetime, is a maximally symmetric space-time, but with constant positive curvature. Analysing quantum scalar fields on de Sitter space-time is an old subject \cite{Chernikov:1968zm} and a lot of effort has been spent in studying it since then  \cite{ Ford:1984hs, Allen:1985ux, Antoniadis:1985pj, Allen:1987tz, Polarski:1991ek, Kirsten:1993ug, Ratra:1984yq, Dolgov:1994ra, Takook:2000qn, Tolley:2001gg, Garbrecht:2006df, Page:2012fn,  Tanaka:2013caa, Woodard:2004ut, Miao:2010vs, Akhmedov:2013vka, Wetterich:2015gya, Wetterich:2015iea}). Like the flat spacetime, all the quantum information of  free fields on classical de Sitter geometry, gets encoded in the Wightman function. From this Wightman function one can further construct the RHS of the Einstein's field equations, i.e., the vacuum expectation of the stress-energy operator \cite{ Parker:2009uva}. Like all quadratic operators in quantum field theory, the  Wightman function itself becomes ill-defined  at the same space-time point, since for well behaved states, it has the so-called Hadamard form i.e., diverges quadratically as well as logarithmically \cite{Parker:2009uva}. Such divergences are attributed to the ultraviolet limit of the theory and they typically get regularized in most of the physical scenarios. However, if we consider the Wightman function of a  minimally coupled massless scalar field in de Sitter universe,  it shows up divergence even for different space-time points, a phenomenon known as infra-red problem of the de Sitter space-time. This infrared divergence of the Wightman function is also intimately tied up with the prediction of scale invariance of the power spectrum \cite{Parker:2009uva, Lochan:2018pzs} which is one of the remarkable success story of the inflationary paradigm. However, this divergence also goes on to suggest that for de Sitter (or de Sitter like universes) the quantum fluctuations can become very important. Various physical reasons of these divergences are suggested in literature and in order to obtain physically meaningful results from the Wightman function, various methods have been devised over the years (\cite{Dowker:1975tf,Parker:2009uva}).  Still the infrared problem calls the stability of de-Sitter spacetime into question, as the severity of these divergences may also be felt up at loop levels in quantum field theory \cite{Akhmedov:2012dn}. In \cite{Lochan:2018pzs} quantum fields over  a family of FRW universes were shown to be connected to  quantum fields in de Sitter spacetime and hence sharing the divergences as well, in some cases. Thus, the potential instability of the de-Sitter spacetime may also have adverse effects for the stability of these Friedmann universes. However, if we wish to study the stability of  such spacetimes through the semiclassical Einstein's equations, divergence in Wightman function may not be sufficient or reliable enough. In order to investigate if the quantum  fluctuations are strong enough to make the de Sitter or the connected  FRW universes unstable under stochastic gravity approach, we need to evaluate the relevant noise kernel components too.

 The noise kernel can also be expressed in terms of the products of derivatives of Wightman functions \cite{Frob:2013sxa}  and therefore the divergences of Wightman functions may also creep in the expressions for the noise kernel. In this paper, we evaluate the noise kernel,  for  minimally coupled massive scalar field in de Sitter universe and for the connected massless fields in Friedmann space-time. We obtain the noise kernel in the late time universe when the scale factor grows and the physical distance between fixed co-moving co-ordinates becomes large. We analyze if the quantum fluctuations which were non-zero initialy over small physical distances, retain their form, grow or decay, as the scale factor growth seperates the points apart.  However, the stress energy tensor, being quadratic in nature has an in-built ultra-violet divergence in it, the noise kernel is obtained from the so called Regularized Stress Energy Tensor (RSET).  For the flat Minkowski space-time, \cite{Frob:2013sxa} calculates the noise kernel for a massive scalar field and gives a dimensional regularization procedure to separate the problematic parts.

 In this paper, we adopt the formalism discussed in \cite{Frob:2013sxa} but generalized for Friedmann universes. Study of quantum fields and their back reaction has attracted a lot of attention  \cite{Parker:1968mv, Parker:1969au, Parker:1971pt, Mottola:1984ar, Fulling:1989nb, Parker:1999td, Hollands:2014eia}. However, in this work, we are interested in quantifying the stochasticity in Einstein Langevin approach for such spacetimes. To begin with, we calcualte the noise kernel for a massive scalar field  in the de Sitter spacetime and analyse the stability of it.  A  similar study has been done in \cite{PerezNadal:2009hr} which  suggests the decay of stochastic noise over large distances,  as the mass of the field increases. However, it is important to note that, in FRW universes the spacelike distances can become large in two ways --(i) For large comoving distances and non-zero scale factor, or--(ii) fixed finite co-moving distances but with large scale factor. In flat spacetime, there are non-zero quantum fluctuations for finite distances and as distances grow, the fluctuations decay and become sub-dominant in front of any other relevant expectation values. However, in certain Friedmann universes, it may so happen that the finite co-moving fluctuations remain invariant or even grow when the scale factor rises and makes the physical distances large. The first may be expected for conformal field theories, something which we see in de Sitter universes too. However, we show that there exists a class of theories in de Sitter universe, where noise kernel blows up as the physical distance between fixed comoving points grows. Subsequently, through the relation to Friedmann universes, we realize  that there are universes, where the stochastic correction in semiclassical analysis should really become important.  We show that the universe driven by pressureless dust or strong energy conditions violating fluids (accelerating universes belong to family of such solutions), becomes very susceptible to such fluctuations and the semiclassical understanding or stabilities of these spacetimes, may need to be reinvestigated in face of divergent fluctuations.  Further, there are other phantom fluid driven Friedmann universes  which retain the initial quantum fluctuations under scale factor growth (much like conformal field theories, despite not being one) and  there the semiclassical analysis should be weighed against the strength of the quantum fluctuations. We also do the analysis for non-minimal coupling and show that a conformal interaction is able to cure this blow up in all cases.
 

The paper has five sections. In section \ref{Prelim}, we quickly review some standard results regarding de Sitter space-time and quantization of a minimally coupled scalar field living on classical de Sitter space-time as well as a brief discussion of evaluation of noise kernel. In section \ref{deSitterNK}, we derive the expression of the noise kernel for a minimally coupled massive scalar field living on de Sitter universe and analyse its various mass limits. We develop the noise kernel computation both for minimal and non-minimal interaction case. Section \ref{FRWNK} deals with space-times which are conformally dual to de Sitter spacetimes in terms of quantum field analysis. We compute the noise kernel in these cases and study the stability here. Section \ref{EnergyCorr} deals with energy density correlator and  the subsequent analysis for Friedmann universes  driven by various matter equation of states.  We summarise our main results and discuss future prospects in conclusion in Section \ref{Summary}. We use the (-, +, +, +) sign convention for the metric.

\section{Preliminaries} \label{Prelim}
\subsection{Noise Kernel}
In the Stochastic gravity paradigm, one tries to include the effect of quantum fluctuations of the matter field on the classical geometry of the space-time through noise kernel, which is given by \cite{Phillips:2000bp, Hu:2008rga} :
\bea\label{eq15}
N_{abcd} &=& \frac{1}{8} \{\hat{t}_{ab}, \hat{t}_{cd} \}, \text {where, }  \hat{t}_{ab}  =
\hat{T}_{ab} -\langle\hat{T}_{ab} \rangle, \text{ and,} \\
 \langle \hat{t}_{abcd}(x,x') \rangle &\equiv & \langle \hat{t}_{ab}(x)\hat{t}_{cd}(x') \rangle=\langle 0|\hat{T}_{ab}(x)\hat{T}_{cd}(x')|0 \rangle - \langle 0|\hat{T}_{ab}(x)|0 \rangle\langle 0|\hat{T}_{cd}(x')|0 \rangle \, .
\eea 
Here $\hat{T}_{ab}$ represents the stress-energy (or energy-momentum) quantum operator which one obtains by replacing the classical fields by field operators in the classical expression for the stress-energy tensor. The noise kernel incorporates many information regarding the quantrum property of matter and its back reaction on the geometry too. For example, one can easily obtain the fluctuation in stress energy tensor from the noise kernel. Fluctuation in stress energy tensor at some space-time point $x$, like any other quantum operator, is equal to $\langle(\hat{t}_{ab}(x))^2\rangle$. Therefore, we see that the fluctuation in stress energy tensor is obtained by taking $a = c$ and $b = d $ and by considering the $x' \to x$ limit in the noise kernel i.e., $\underset{x'\to x}{\lim} \langle \hat{t}_{abab}(x,x') \rangle$ \footnote{Fluctuations obtained in this way are generally divergent in the $x' \to x$ limit, but these can be taken care of, by proper regularization procedures as is routinely  done for divergent observables in quantum field theory \cite{Phillips:2000bp, Moretti:1997qn, Parker:2009uva}. However, in the present paper, we are more interested in the correlations rather than the fluctuations explicitly.}.
\\ Further, in gravitational scenario, stress-energy tensor, denoted by $T_{\alpha\beta}$,  is defined as the variation of the matter action with respect to the metric variation i.e.,
\begin{equation}
T_{\alpha\beta}(x) = -\frac{2}{\sqrt{-g}}\fdv{S_{M}}{g^{\alpha \beta}(x)}\,.
\end{equation}
Therefore, for a minimally coupled massive scalar field in general space-time metric, given by
\begin{equation}\label{eq17}
S[g_{\alpha\beta},\phi]= -\frac{1}{2}\int d\eta d^{3}\vec{x} \,\sqrt{-g}\, \big( g^{\alpha \beta}\,\nabla_{\alpha}\phi \,\nabla_{\beta} \phi  + m^2\phi^2\big)\, ,
\end{equation}
we have 
\begin{equation}
T_{\alpha\beta}(x) = \nabla_{\alpha}\phi \nabla_{\beta}\phi - \frac{1}{2}g_{\alpha\beta}(g^{\gamma\delta}\nabla_{\gamma}\phi \nabla_{\delta}\phi + m^2 \phi^2)\, .
\end{equation}
For example,  if we consider the Minkowski space i.e.,  $g_{\alpha \beta} = \eta_{\alpha\beta} $, we have 
\begin{equation} \label{eq18}
T_{ab}(x) = \underset{y\to x}{\lim} P_{ab}(x,y)\phi(x)\phi(y) \, , 
\end{equation}
where 
\begin{equation}
P_{ab}(x,y) =  \big(\delta_{(a}^{c}\delta_{b)}^{d}-\frac{1}{2}\eta_{ab}\eta^{cd}\big)\nabla_{c}^{x}\nabla_{d}^{y} - \frac{1}{2}\eta_{ab} m^2 \, .
\end{equation}

\noindent This implies that the stress-energy two point correlator in the vacuum of the field can be obtained as \footnote{In this expression and the following expressions, we don't bother ourselves to put hat over the stress-energy (or field) operators since it is understood that we are exclusively working with quantum stress-energy operators. } :
\begin{multline}
 \langle t_{abcd}(x,x')\rangle=
 \underset{\underset{y' \to x'}{y\to x}} {\lim} P_{ab}(x,y)P_{cd}(x',y') \langle 0|\phi(x)\phi(y)\phi(x')\phi(y')|0 \rangle \\ -  \underset{\underset{y' \to x'}{y\to x}} {\lim}P_{ab}(x,y)P_{cd}(x',y')\langle 0|\phi(x)\phi(y)|0 \rangle\langle 0|\phi(x')\phi(y')|0 \rangle  \, . 
\end{multline}
After some manipulations, this becomes \cite{Frob:2013sxa}
\begin{equation}
\langle t_{abcd}(x,x')\rangle = 2 \underset{\underset{y' \to x'}{y\to x}} {\lim}  P_{ab}(x,y)P_{cd}(x',y') G(x,x')G(y,y')\, ,  \label{correlator}
\end{equation}
where $G(x,x')$ is the Wightman function for the scalar field in the considered vacuum, 
\begin{equation}
G(x,x') = \bra{0}\phi(x)\phi(x') \ket{0} \, .
\end{equation}

\subsection{de Sitter Space}
An n-dimensional de Sitter space, denoted by $dS_{n}$, can be viewed as the embedding 
\begin{equation}
\eta_{ab}X^{a}X^{b} = H^{-2},
\end{equation} 
in $\mathbb{R}^{(1,n)}$ with the metric $\eta_{\mu \nu} = diag(-1, 1,....., 1)$. The de Sitter space is a maximally symmetric space and has constant Ricci Scalar, $R = 2 (n-1)(n-2)H^2$. It can be shown that the de Sitter space is also the solution of vacuum Einstein equations with a positive cosmological constant, given by $(n-1)(n-2)H^2/2$.

One can use a number of coordinate systems to cover the de Sitter space\footnote{To know more about different coordinate systems used for de Sitter space (like Global coordinates, Static coordinates, Eddington-Finkelstein coordinates, Kruskal coordinates  etc.) and the causal structure of the de Sitter space (i.e., its Penrose diagram etc.), one can refer to \cite{Hawking:1973uf},\citep{Kim:2002uz},\cite{Spradlin:2001pw}.} but one particularly useful coordinate system (called planar or inflationary coordinates) for our purposes is given by :
\begin{eqnarray}
X^n - X^0 & = & \pm \frac{e^{Ht}}{H}, 
X^i  =  \pm x^i e^{Ht}, i = 1,....., n-1, \nonumber\\
X^n + X^0 & = &\pm \Big(\frac{e^{-Ht}}{H} -  x_i x^i H e^{Ht}\Big) .
\end{eqnarray}
For a given sign in the above equation one covers only half of the de Sitter manifold. Therefore, $\pm$ signs correspond to two charts covering the full de Sitter space. In a single chart, all the coordinates (i.e., $t, x^1,x^2,....,x^{n-1}$) lie between $(-\infty, \infty)$. \\
In these coordinates, the metric in both the charts is given by:
\begin{equation}
ds^2 = -dt^2 + e^{2Ht}d\vec{x}^2 \, .
\end{equation}
The transformation\footnote{Now $\eta$ lies between $(-\infty, 0)$ corresponding to $t$ lying between $(-\infty, \infty)$.} $d\eta = dt/a(t)$ with $a(\eta) = -1/H\eta$ brings the above metric in the following conformally flat form :
\begin{equation}
ds^2 = \frac{1}{(H\eta)^2}(-d\eta^2 + d\vec{x}^2) \, .
\end{equation}
If we define $Z(x,x') = H^2\eta_{ab}X^a(x)X^b(x')$, then in the planar coordinates, we have 
\begin{equation}\label{eq6}
Z(x,x') = 1 + \frac{(\eta -\eta')^2 - (\vec{x} -\vec{x}')^2}{2\eta\eta'} \, . 
\end{equation}
This is a useful quantity, which (indirectly) characterizes the geodesic distance between points $x$ and $x'$ on the de Sitter manifold.

\subsection{Quantum Fields on de Sitter Space}
The minimally coupled scalar field, corresponding to (\ref{eq17}), satisfies the following equation of motion: 
\begin{equation}
(\Box - m^2)\phi(x) = 0 \, .  
\end{equation}
The Wightman function also satisfies the same equation as the field, i.e.,
\begin{equation}\label{eq1}
 (\Box - m^2)G(x,x') = 0 \, .
\end{equation}
For a de-Sitter invariant vacuum, evidently this depends only on the geodesic distance, i.e., $ G(x,y) = G(Z(x,y))$  and the above equation becomes (see \cite{Allen:1985ux},\citep{Allen:1985wd}):
\begin{equation}
(Z^2 -1)\dv[2] {G}{Z} + 4Z\dv{G}{Z} + \frac{m^2}{H^2}G(Z) =0 \, .
\end{equation}
Under the transformation $Z \to Y =  (1+Z)/2$, it further reduces to :
\begin{equation}
Y(1-Y)\dv[2] {G}{Y} + (2 - (a + b + 1) Y)\dv {G}{Y} - a b G(Y) = 0
\end{equation}
where $a = 3/2 + \sqrt{9/4 - m^2/H^2}$ and $b = 3/2 - \sqrt{9/4 - m^2/H^2}$ or vice-versa. This is the hypergeometric equation and one particular solution to this equation is $G(Z) = {_2}F_1(a,b,2,\frac{1+Z}{2})$. A particular choice for the de Sitter invariant vacuum state (called the Bunch-Davies Vacuum) leads to $G(Z) = (H^2/16\pi^2)\Gamma(a)\Gamma(b) _2F_1(a,b,2,\frac{1+Z}{2}) $.
Due to its structure, massless fields (for which $a=3,b=0$) have a divergent piece in $G(Z)$ which is identified as the infrared divergence. Since the Wightman function is only a function of $Z$, all its higher order derivatives can be evaluated in terms of variation of $Z$ w.r.t. space-time co-ordinates (see Appendix \ref{CorrDerr}).


\section{Noise Kernel in de Sitter universe} \label{deSitterNK}
First we express the stress energy correlator (algebraically related to the noise kernel) in a conformally flat Friedmann space-time. For a minimally coupled scalar field in a conformally flat space-time metric,  i.e.,  $g_{\alpha \beta} = a(\eta)^2 \eta_{\alpha\beta} $, we can again use the expression (\ref{eq18}) for stress energy correlator but now with 
\begin{equation}
P_{ab}(x,y) =  \big(\delta_{(a}^{c}\delta_{b)}^{d}-\frac{1}{2}\eta_{ab}\eta^{cd}\big)\nabla_{c}^{x}\nabla_{d}^{y} - \frac{1}{2}\Big(\frac{a(\eta) + a(\eta')}{2}\Big)^2\eta_{ab} m^2 \, .
\end{equation}
Now using this expression of $P_{ab}(x,y)$ in equation (\ref{correlator}) and specializing to the case of de Sitter space-time i.e., $a(\eta) = -\frac{1}{H\eta}$, we have the following expression for noise kernel:
\begin{multline}\label{eq5}
\langle t_{abcd}(x,x')\rangle_{dS} = \Bigg(\nabla_{b}\nabla_{c}'G(x,x')\nabla_{a}\nabla_{d}'G(x,x') + \nabla_{b}\nabla_{d}'G(x,x')\nabla_{a} \nabla_{c}'G(x,x') \\- \eta_{cd}\eta^{\rho \sigma}\nabla_{a}\nabla_{\rho}'G(x,x')\nabla_{b}\nabla_{\sigma}'G(x,x') - \frac{1}{H^2\eta'^2}m^2 \eta_{cd}\nabla_{a}G(x,x')\nabla_{b}G(x,x') \\ -\eta_{ab}\eta^{\gamma \delta}\nabla_{\gamma}\nabla_{c}'G(x,x')\nabla_{\delta}\nabla_{d}'G(x,x') +  \frac{1}{2}\eta_{ab}\eta^{\gamma \delta}\eta_{cd}\eta^{\rho \sigma}\nabla_{\gamma}\nabla_{\rho}'G(x,x')\nabla_{\delta}\nabla_{\sigma}'G(x,x') \\ 
 + \frac{1}{2H^2\eta'^2}m^2\eta_{ab}\eta^{\gamma \delta}\eta_{cd}\nabla_{\gamma}G(x,x')\nabla_{\delta}G(x,x') - \frac{1}{H^2\eta^2}m^2 \eta_{ab}\nabla_{c}'G(x,x')\nabla_{d}'G(x,x') \\ + \frac{1}{2H^2\eta^2}m^2 \eta_{ab}\eta_{cd}\eta^{\rho \sigma}\nabla_{\rho}'G(x,x')\nabla_{\sigma}'G(x,x') + \frac{1}{2H^4\eta^2\eta'^2}m^4 \eta_{ab}\eta_{cd}G(x,x')G(x,x')\Bigg) \, .
\end{multline}
We are interested in learning if the primordial fluctuations remain relevant if the universe expands. For this purpose, we first choose a space-like surface by fixing $\eta$. We now use the properties of the Wightman function on constant time ($\eta-$) hypersurfaces and evaluate $\langle t_{abcd}(x,x')\rangle_{dS}$ when the physical distances between fixed co-moving distances grow very large, i.e., $a(\eta) \rightarrow \infty$, which in expanding universes will be the late time era.

\subsection*{Minimal coupling}
In order to study the stochastic correction, in principle, it will be necessary to consider all the components of the noise kernel. However,  for our purpose, it will be sufficient to explore only the $\langle t_{0000}\rangle$ component to establish the growth or decay of such stochastic corrections.  In fact, the table in the Appendix \ref{PowerCount} shows that the degree of divergence (if any) of the other components of the noise kernel is either less than or equal to that of the $\langle t_{0000}\rangle$ component of the noise kernel. Further the $\langle t_{0000}\rangle$ also gives the energy energy correlator in a straight forward manner which is a readily accessible observable quantity \cite{Akrami:2018odb}.  
Therefore, we need to calculate the $(a =0, b=0, c=0, d=0)$ component of the noise kernel. In de Sitter space-time, late-time corresponds to $\eta \to 0$ limit. So, we consider the noise kernel on constant time sheets (i.e., $\eta = \eta'$) with finite spatial distances (i.e., $ \Delta \vec{x} \neq 0$) and then we take the $\eta \to 0$ limit.\\
Using the equation (\ref{eq5}) and formulae from Appendix \ref{CorrDerr}, we see that 
\begin{multline}\label{eq7}
\langle t_{00}(\eta,\vec{x})t_{00}(\eta, \vec{x'})\rangle_{dS} = \Bigg((G'')^2\Big[\frac{(\Delta \vec{x})^6}{4 \eta^{10}} + \frac{(\Delta \vec{x})^8}{32\eta^{12}} + \frac{(\Delta \vec{x})^4}{2\eta^{8}}\Big]  + G^2 \Big [\frac{m^4}{2H^4\eta^4}\Big]\\ + (G')^2\Big[\frac{3(\Delta \vec{x})^2}{2\eta^6}  + \frac{(\Delta \vec{x})^4}{8\eta^8} + \frac{2}{\eta^4} + \frac{m^2}{H^2}\Big(\frac{(\Delta \vec{x})^4}{4\eta^8} + \frac{(\Delta \vec{x})^2}{\eta^6}\Big)\Big] + (G''G')\Big[-\frac{5(\Delta \vec{x})^4}{4\eta^8} - \frac{(\Delta \vec{x})^2}{\eta^6} - \frac{(\Delta \vec{x})^6}{8\eta^{10}}\Big]\Bigg) \, .
\end{multline}
Using the expressions for the Wightman function and its derivatives in the Bunch Davies Vacuum, i.e., 
\begin{eqnarray}
G(Z) &=& \frac{H^2}{16\pi^2}\Gamma\Big(\frac{3}{2} + \nu\Big)\Gamma\Big(\frac{3}{2} - \nu\Big) {_2}F_1\Big(\frac{3}{2} + \nu,\frac{3}{2} - \nu,2,\frac{1+Z}{2}\Big) \, , \label{eq8} \\
G'(Z) &=& \frac{H^2}{64\pi^2}\Gamma\Big(\frac{5}{2} + \nu\Big)\Gamma\Big(\frac{5}{2} - \nu\Big) {_2}F_1\Big(\frac{5}{2} + \nu,\frac{5}{2} - \nu,3,\frac{1+Z}{2}\Big) \, , \label{eq9}  \\
G''(Z) &=& \frac{H^2}{384\pi^2}\Gamma\Big(\frac{7}{2} + \nu\Big)\Gamma\Big(\frac{7}{2} - \nu\Big) {_2}F_1\Big(\frac{7}{2} + \nu,\frac{7}{2} - \nu,4,\frac{1+Z}{2}\Big)  \, , \label{eq10} 
\end{eqnarray}
\Big(where $\nu = \sqrt{\frac{9}{4}-\frac{m^2}{H^2}}$\Big), and appealing to the late time-behaviour (i.e., $ Z \to -\infty$) for the $_2F_1$ functions \citep{abramowitz+stegun} i.e., 
\begin{multline}
_2F_1(a,b,c,z) = \frac{\Gamma(b-a)\Gamma(c)(-z)^{-a}}{\Gamma(b)\Gamma(c-a)}\Big(\sum_{k=0}^{\infty}\frac{(a)_k(a-c+1)_kz^{-k}}{k!(a-b+1)_k}\Big) \\ \\ + \frac{\Gamma(a-b)\Gamma(c)(-z)^{-b}}{\Gamma(a)\Gamma(c-b)}\Big(\sum_{k=0}^{\infty}\frac{(b)_k(b-c+1)_kz^{-k}}{k!(b-a+1)_k}\Big) \, ,
\end{multline}  
we have
\begin{multline}
\langle t_{00}(\eta,\vec{x})t_{00}(\eta, \vec{x}')\rangle_{dS}\big|_{\text{late time}}  = \frac{H^4\Gamma^2(\nu)\Gamma^2(\frac{5}{2}-\nu)}{\pi^5}\Bigg[\frac{9 \eta^{2-4\nu}}{32(\Delta \vec{x})^{6-4\nu}} \\ \\  + \frac{21 (3-2\nu) \eta^{4-4\nu}}{16(\Delta \vec{x})^{8-4\nu}} + \frac{(656 \nu^3 -3244 \nu^2 + 5168 \nu - 2655)\eta^{6-4\nu}}{64 (\nu-1)(\Delta \vec{x})^{10-4\nu}} + O(\eta^2)\Bigg].
\end{multline}

From here, we can see that there is a transition in the behaviour of the stochastic correction term at $\nu =1/2$. For any $\nu <1/2$, it vanishes in the late lime limit (large physical distances) as $\eta^{2-4\nu}$, e.g., for $\nu = 0$, 
\bea
\langle t_{00}(\eta,\vec{x})t_{00}(\eta, \vec{x}') \rangle_{dS}\big|_{\text{late time}}  = 
 \lim_{\eta \to 0} \Bigg[ O (\eta) \Bigg].
\eea
On the other hand, the noise kernel approaches a saturating value at $\nu = 1/2$, for large physical distances with finite co-moving distance,
\bea
\langle t_{00}(\eta,\vec{x})t_{00}(\eta, \vec{x}') \rangle_{dS}\big|_{\text{late time}}  =  \lim_{\eta \to 0} \Bigg[\frac{9H^4}{32\pi^4(\Delta \vec{x})^4} + O(\eta)  \Bigg].
\eea
This is not surprising as $\nu = 1/2$ is conformal field theory and does not feel  $a(\eta)$. However, this goes on to suggest that the stochastic correction is uncontrollable after $\nu >1/2$.
For example, for a massless field, $\nu = 3/2$,
\begin{multline}
\langle t_{00}(\eta,\vec{x})t_{00}(\eta, \vec{x}') \rangle_{dS} \big|_{\text{late time}} =\\  \lim_{\eta \to 0}\lim_{\epsilon \to 0 } \Bigg[ \frac{9H^4}{128\pi^4 \eta^4}\Big[1 - 4 \epsilon\Big]  + \frac{21H^4 \epsilon}{32\pi^4(\Delta \vec{x})^2\eta^2} + \frac{H^4}{16\pi^4(\Delta \vec{x})^4}\Big[\frac{3}{2} + 14 \epsilon\Big] + O (\eta) \Bigg] \\ =   
  \lim_{\eta \to 0} \Bigg[ \frac{9 H^4}{128 \pi^4 \eta^4} + \frac{H^4}{16\pi^4(\Delta \vec{x})^4}\Big[\frac{3}{2} \Big] + O (\eta) \Bigg] \rightarrow \infty.
\end{multline}

Therefore, we see that, for a minimally coupled scalar field in Bunch Davies Vacuum, the $(a=0, b=0, c=0, d=0)$ component of the noise kernel (on constant time sheets with finite spatial distance and in the late time universe limit) undergoes a kind of 'phase transition' as a function of $\nu$, with critical value being $\nu = 1/2$.  To put in context, it is also well known that de Sitter has an instability against the particle creation of light mass particles \cite{ Ford:1984hs, Krotov:2010ma, Anderson:2013ila, Anderson:2013zia, Anderson:2017hts}.

\subsection*{Comparison with large co-moving distance case}
At this point, we can compare our results with the case for large co-moving distance case, obtained in \cite{PerezNadal:2009hr} in which the noise kernel is shown to be: 
\begin{multline}
\langle t_{abcd}(x,x')) \rangle_{dS} = P(\mu)n_an_bn_cn_d + Q(\mu)(n_an_bg_{c'd'} + n_{c'}n_{d'}g_{ab} )\\ \\ + R(\mu)( n_{s}n_{c'}g_{bd'} +  n_{b}n_{d'}g_{ac'} +  n_{a}n_{d'}g_{bc'} +  n_{b}n_{c'}g_{ad'} )  + S(\mu)(g_{ac'}g_{bd'} + g_{bc'}g_{ad'}) + T(\mu)g_{ab}g_{c'd'} \, ,
\end{multline}
where $P,Q,R,S,T$ are sums of products of Wightman function and its first and 2nd order derivatives with respect to the geodesic distance. Here, $n_a$ and $n_{a'}$ are the unit tangent vectors to the geodesic connecting the points $x$ and $x'$, at $x$ and $x'$, respectively. The action of $g_{ac'}$ is to parallel-transport a vector from $x'$ to $x$ along the geodesic. \\
For $Z<< -1$  regime, $P, Q, T \sim Z^{-2h\_ }$ and  $R \sim Z^{- 2 h\_ -1}$  and $S \sim  Z^{-2h\_ -2}$. Using these behaviours of $P,Q,R,S$ and $T$, it is argued that \textit{the fluctuations decay faster with the distance as mass increases}. However, for the fixed co-moving distance and large scale factor limit,  the coefficients of $P(\mu), Q(\mu)$ etc., in the above equation, also depend upon $\eta$ \big(and hence on $Z$ as $Z = 1 + \frac{(\eta -\eta')^2-(\Delta \vec{x})^2 }{2\eta\eta'}$\big) and the mentioned result is obtained ignoring these dependences. So, in this sense, the results of \cite{PerezNadal:2009hr} are, in fact, valid for those scenarios in which $\eta$ and $\eta'$ are held finite (and constant) and $Z$ approaches large values through $(\Delta \vec{x})^2 \to \infty$ limit. However, large spatial separation can arise in another way, namely with finite $\Delta \vec{x} (\neq 0)$ and $a(\eta) \to \infty$. This other scenario again shows up the divergences obtained in previous section, for the relevant mass ranges.

A similar expression can be derived for non-minimally coupled fields as well. Though the relations derived above carry over with simple reparameterization $m^2 \rightarrow  m^2 + 12 \xi H^2$, we still present a brief discussion for the non-minimal case.
\subsection*{Non-minimal coupling}
Let's consider a non-minimally coupled massive scalar field with the following action\footnote{Here superscript nm refers to non-minimal coupling.}
\begin{equation}
S^{nm}[g_{\alpha\beta},\phi]= -\frac{1}{2}\int d\eta d^{3}\vec{x} \,\sqrt{-g}\, \big( g^{\alpha \beta}\,\nabla_{\alpha}\phi \,\nabla_{\beta} \phi + m^2\phi^2 + \xi R\phi^2\big)\, .
\end{equation}
It gives the following equation of motion for the scalar field, $\phi$:
\begin{equation}
\big[\Box - (12\xi H^2 + m^2)\big]\phi(x) = 0\, ,
\end{equation}
which implies
\begin{equation}
G(Z(x,x')) = \frac{H^2}{16\pi^2}\Gamma(a)\Gamma(b) _2F_1\big(a,b,2,\frac{1+Z}{2}\big) \,  ,
\end{equation}
where $ a = \frac{3}{2} + \sqrt{\frac{9}{4} - \frac{12\xi H^2 + m^2}{H^2}}$ and $b = \frac{3}{2} - \sqrt{\frac{9}{4} - \frac{12\xi H^2 + m^2}{H^2}}$. \\ \\
The stress-energy tensor for this case is given by:
\begin{multline}
T^{nm}_{\alpha\beta}(x) = \nabla_{\alpha}\phi \nabla_{\beta}\phi - \frac{1}{2}g_{\alpha\beta}(g^{\gamma\delta}\nabla_{\gamma}\phi \nabla_{\delta}\phi + m^2 \phi^2)  + \xi \big(G_{\alpha\beta}\phi^2 + g_{\alpha\beta} g^{\gamma\delta}\nabla_{\gamma} \nabla_{\delta}\phi^2 - \nabla_{\alpha}\nabla_{\beta}\phi^2\ \big) \, ,
\end{multline}
where $G_{\alpha\beta}$ is the Einstein tensor.
Using the fact that, for de Sitter space, $G_{\alpha\beta} = -3H^2 g_{\alpha\beta}$ and $g_{\alpha\beta} = \eta_{\alpha\beta}/H^2\eta^2$, we have
\begin{equation}\label{eq16}
T^{nm}_{\alpha\beta}(x) = \underset{{y\to x}} {\lim}P^{nm}_{ab}(x,y)\phi(x)\phi(y)  = \underset{{y\to x}} {\lim}\big(P_{ab}(x,y) + M_{ab}(x,y) \big)\phi(x)\phi(y) \, ,
\end{equation}
where 
\begin{equation}\label{eq12}
P_{ab}(x,y)  = \Bigg[ \Big((1-2\xi)\delta_{(a}^{r}\delta_{b)}^{s}-(\frac{1}{2}-2\xi)\eta_{ab}\eta^{rs}\Big)\nabla_{r}^{x}\nabla_{s}^{y} - \frac{2(3H^2\xi + \frac{m^2}{2})}{(H\eta)^2 + (H\eta')^2}\eta_{ab} \Bigg]  \, , 
\end{equation}
and 
\begin{equation}\label{eq13}
M_{ab}(x,y) = \Bigg[2\xi\eta_{ab}\eta^{rs} - 2\xi\delta_{(a}^{r}\delta_{b)}^{s}\Bigg]\frac{\nabla_{r}^{x}\nabla_{s}^{x} + \nabla_{r}^{y}\nabla_{s}^{y}}{2}   \, .
\end{equation}
In the above formula, $\eta$ and $\eta'$ correspond to the time coordinate of points $x$ and $y$, respectively. Here, we see that the $P_{ab}$ part is the same as it is for the minimally coupled scalar field with $\xi = 0$. Also, the expression for $P^{nm}_{ab}(x,y)$ is symmetric in $x$ and $y$. \\ 
\noindent Similar to the minimally coupled case, we find that
\begin{multline}\label{eq14}
\langle t^{nm}_{ab}(x)t^{nm}_{cd}(x') \rangle = 2 \underset{\underset{y' \to x'}{y\to x}} {\lim}  P^{nm}_{ab}(x,y)P^{nm}_{cd}(x',y') G(x,x')G(y,y')  \\  = 2 \underset{\underset{y' \to x'}{y\to x}} {\lim}  \Big(P_{ab}(x,y)P_{cd}(x',y') + P_{ab}(x,y)M_{cd}(x',y') \\ + M_{ab}(x,y)P_{cd}(x',y') + M_{ab}(x,y)M_{cd}(x',y')\Big) G(x,x')G(y,y')   \, .
\end{multline}
The contributions of the $P_{ab}P_{cd}$, $P_{ab}M_{cd}$, $M_{ab}P_{cd}$ and $M_{ab}M_{cd}$ terms, to the noise kernel expression, are given in the Appendix \ref{NonMinimal}. The power counting argument clearly shows that the most dominant power of $\eta$  (in the limit $\eta \to 0$), in the relevant noise kernel component, is still $2-4\nu$ \big(where $\nu = \sqrt{\frac{9}{4} - \frac{12\xi H^2 + m^2}{H^2}}$\big). In fact, we have

\begin{multline}
\langle t^{nm}_{ab}(x)t^{nm}_{cd}(x')\rangle\big|_{\text{late time}} = \underset{\eta \to 0}{\lim}\Bigg[\frac{\eta^{2- 4 \nu}H^4}{512\pi^5 (\Delta \vec{x})^{6-4\nu}}\Big[ 32(12\xi -1)\Gamma\big(\frac{5}{2}-\nu\big)\Gamma\big(\frac{7}{2} - \nu\big) \\ + \Big(16\frac{m^4}{H^4} + 8 \frac{m^2}{H^2}(24 \xi + (3 - 2\nu)^2) - 48\xi(3-2\nu)^2 + (3-2\nu)^2(29-20\nu + 4\nu^2)\\ + 32\xi^2(27-12\nu + 4\nu^2))\Big)\Gamma^2\big(\frac{3}{2}- q\big)\Big]\Gamma[\nu]^2 + O(\eta^{4-4\nu})\Bigg].
\end{multline}
 This implies that the noise kernel for a conformally coupled scalar field behaves exactly similar to the noise kernel for a minimally coupled scalar field except for the fact that $m^2/H^2$ in the latter case goes to $m^2/H^2 + 12\xi$ in the former i.e., it undergoes a sort of ``divergent-transition'' as $m^2/H^2 + 12\xi$ crosses the critical value 2, making $\nu \geq 1/2$. Thus, we readily see that the conformal coupling $\xi =1/6$ cures the divergence as even for the massless field, we get $\nu =1/2$, which, at best, has a non-zero finite value of noise kernel component over large physical scales. For any other non-zero mass, the value of $\nu$ is less than 1/2, showing a vanishing correlation over large scales. However, for any $\xi <1/6$, we still have divergences over a range of mass values. Clearly, this divergence in late times is different from the secular divergences of stress-energy as (a) we use Regularized Stress Energy Tensor (RSET), and (b) this divergence appears only for finite co-moving distance in the large scale factor limit. Thus, the correlation structure on fixed co-moving distance may grow or decay as the scale factor turns large, depending upon the value of the coupling $\xi$ and mass $m$. We now relate the noise kernel of Friedmann universes with the noise kernel of de Sitter universe for various masses. 

\section{Noise Kernel for Friedmann spaces} \label{FRWNK}
In this section, we relate the results of the previous sections on the components of the noise kernel in the de Sitter space-time to the corresponding noise kernel components in Friedmann space-times using the fact that a massless scalar field in a Friedmann space-time is conformally equivalent to a massive scalar field in de Sitter space-time \footnote{The action of a massless scalar field in a universe with metric, $g_{\alpha\beta}= a^2\eta_{\alpha\beta}$ with $a(\eta) = (H\eta)^{-q}$, is given by
\begin{equation*}
S= -\frac{1}{2}\int d^{4}x \, a^4\big(a^{-2}\,\eta^{\alpha\beta}\partial_{\alpha}\phi \,\partial_{\beta} \phi\big)\, .
\end{equation*}
Under the transformation, $\phi(x) = (H\eta)^{-1+q}\, \psi(x)$, the action becomes 
\begin{equation*}
S= -\frac{1}{2}\int d^{4}x \,b^4\big(b^{-2}\,\eta^{\alpha\beta}\partial_{\alpha}\psi \,\partial_{\beta} \psi - m_{eff}^2 \psi^2 \big)\, ,
\end{equation*}
where $b(\eta) = (H\eta)^{-1}$ and $m^2_{eff} = H^2(1 - q)(2 + q)$. Therefore, we see that a massless scalar field in a Friedmann universe with scaling factor, $a(\eta) = (H\eta)^{-q}$, goes to a massive scalar field in a de Sitter universe under the above mentioned transformation. We also see that the transformation relation between the fields i.e., $\phi(x) = (H\eta)^{-1+q}\, \psi(x)$ explains the relation between the Wightman functions in the related space-times i.e., $G^{P.L.}(x,x') = (H\eta)^{q-1}(H\eta')^{q-1}G(x,x')$. A similar kind of correspondence can be established for non-minimal setting as well. For more details, see Appendix A.2 of \cite{Lochan:2018pzs}.}. If a power-law Friedmann universe has scaling factor, $a(\eta) = (H\eta)^{-q}$, then the corresponding massive scalar field in de Sitter space-time has $m^2 = H^2(1-q)(2+q)$. One also gets that the Wightman function in the power-law universe is related to the the Wightman function in de Sitter space-time, $G^{P.L.}(x,x') = (H\eta)^{q-1}(H\eta')^{q-1}G(x,x')$. Using this, we see that 
\begin{align}\label{eq2}
\nabla_{\mu}'G^{P.L.} = (H)^{2q-2}[(q-1)(\eta)^{q-1}(\eta')^{q-2}G \delta_{\mu0} + (\eta)^{q-1}(\eta')^{q-1}\nabla_{\mu}'G],
\end{align}
and 
\begin{align}\label{eq3}
\nabla_{\nu}\nabla_{\mu}'G^{P.L.} &= H^{2q-2}[(q-1)^2(\eta)^{q-2}(\eta')^{q-2}G \delta_{\mu0}\delta_{\nu0} + (q-1)(\eta)^{q-1}(\eta')^{q-2}\delta_{\mu0} \nabla_{\nu}G \nonumber \\
 & +  (q-1)(\eta)^{q-2}(\eta')^{q-1}\delta_{\nu0} \nabla_{\mu}'G + (\eta)^{q-1}(\eta')^{q-1}\nabla_{\nu}\nabla_{\mu}'G] \nonumber \\
&  = (H\eta H\eta')^{q-1}\Big(\frac{(q-1)^2}{\eta\eta'}G \delta_{\mu0}\delta_{\nu0} + \frac{(q-1)}{\eta'}\delta_{\mu0} \nabla_{\nu}G
  +  \frac{(q-1)}{\eta}\delta_{\nu0} \nabla_{\mu}'G +\nabla_{\nu}\nabla_{\mu}'G\Big) \, . 
\end{align} 
The above expression, for $\eta = \eta'$, for different values of $\nu$ and $\mu$ is given by (see Appendix \ref{CorrDerr}):
\bea
 \nabla_{0}\nabla_{0}'G^{P.L.} & =& (H\eta)^{2q-2}\Big[\frac{(q-1)^2}{\eta^2}G +   \frac{(q-1)}{\eta}\nabla_{0}G + \frac{(q-1)}{\eta}\nabla_{0}'G + \nabla_{0}\nabla_{0}'G\Big];\nonumber\\
\nabla_{0}\nabla_{j}'G^{P.L.} & =&  (H\eta)^{2q-2}\Big[\frac{q-1}{\eta}\nabla_{j}'G + \nabla_{0}\nabla_{j}'G\Big];\nonumber\\
\nabla_{i}\nabla_{0}'G^{P.L.} & =&  (H\eta)^{2q-2}\Big[\frac{q-1}{\eta}\nabla_{i}G + \nabla_{i}\nabla_{0}'G\Big]; \nonumber\\
\nabla_{i}\nabla_{j}'G^{P.L.} & =&  (H\eta)^{2q-2}\Big[\nabla_{i}\nabla_{j}'G \Big].
\eea

Now, we have the covariant derivatives of Wightman function in the Friedmann space-times in terms of the corresponding quantities in the de Sitter space-time. Using the above expressions in the noise kernel expression for a massless scalar field and for Friedmann space-times, we see that the considered noise kernel component (on constant time sheets) is given by :
\begin{multline}\label{eq4}
\langle t_{00}(\eta,\vec{x})t_{00}(\eta,\vec{x'}) \rangle_{P.L.} = (H\eta)^{4(q-1)}\Bigg[ \frac{G^2}{2\eta^4}(q-1)^4 \  + GG'\Big[\frac{(2q^3 -7q^2 + 8q -3)(\Delta x)^2}{2\eta^6} - \frac{(q-1)^2}{\eta^4}\Big] \\ \\ + GG''\frac{(q-1)^2(\Delta \vec{x})^4}{4\eta^8}  + G'G''\Big[\frac{(q-\frac{3}{2})(\Delta \vec{x})^6}{4\eta^{10}} + \frac{(q-\frac{9}{4})(\Delta \vec{x})^4}{\eta^8} - \frac{(\Delta \vec{x})^2}{\eta^6}\Big]  \\ \\ +  (G')^2\Big[\frac{2}{\eta^4} + \frac{(q^2 - 5q + \frac{11}{2})(\Delta \vec{x})^2}{\eta^6}  + \frac{(2q^2 -6q + \frac{9}{2})(\Delta \vec{x})^4}{4\eta^8}\Big] + (G'')^2\Big[\frac{(\Delta \vec{x})^6}{4\eta^{10}} + \frac{(\Delta \vec{x})^8}{32\eta^{12}} + \frac{(\Delta \vec{x})^4}{2\eta^{8}}\Big]\Bigg]\, . 
\end{multline}
For different power-law universes, i.e., for different values of $q$, one can evaluate the above expression on constant time sheets. However, as $\nu$ can take values only in the range $[-3/2,3/2]$, we see that we can use the considered equivalence only for those values of q which lie in the range $[-2,1]$. The region $|\nu|> 3/2$ is mapped to the region outside $[-2,1]$. As we are interested in the behaviour of the noise kernel component in the late time universe, we observe that, for $q \in (0,1] $, the late time universe corresponds to $\eta \to 0$ and for $q \in [-2,0)$, the late time universe corresponds to $\eta \to \infty$. We now list down the stress-energy correlator for various Friedmann space-times:
\begin{itemize}
\item  $q = 1$ : 
This case trivially corresponds to a massless scalar field in de Sitter space-time, which is just the case $\nu = 3/2$ in the previous section. As discussed above, the correlator diverges in the late-time limit as $\eta^{-4}$ or $a^4$. 

\item $ q \in (0,1)$ : 
If we perform, for this case as well, the same power counting analysis as is done in Appendix \ref{PowerCount}, we find that the relevant noise kernel component in the late time universe i.e., $\eta \to 0$ limit, has an $\eta$ independent term. Therefore, we have a constant late time noise kernel component for those Friedmann universes which have negative exponent of $\eta$  in the scale factor. In fact, we have 
\begin{multline}
\langle t_{00}(\eta,\vec{x})t_{00}(\eta,\vec{x'}) \rangle_{P.L.} \big|_{\text{late time}}= \\\lim_{\eta \to 0} \frac{(H\eta)^{4q-4}}{(\Delta \vec{x})^4}\Bigg[  \frac{H^4\eta^{4-4q}(\Delta \vec{x})^{4q-4}}{8\pi^5}\Bigg((11 - 12q + 4q^2)(\Gamma(2-q))^2(\Gamma(0.5 +q))^2\Bigg)  \\ +\frac{4^{4q}\eta^{4q+4}H^4}{32\pi^5(\Delta \vec{x})^{4+4q}}\Bigg((1+2q)^4(\Gamma(2+q))^2(\Gamma(-0.5-q))^2\Bigg)  + O(\eta^{6-4q})\Bigg] .
\end{multline}
In the late time limit, only the term $\frac{H^{4q}(\Delta \vec{x})^{4q-8}}{8\pi^5}\Big((11 - 12q + 4q^2)(\Gamma(2-q))^2(\Gamma(0.5 +q))^2\Big)$ survives, which is time independent and is therefore remnant of the quantum fluctuations the universe was born with.
For these space-times, the stochastic term, in the Einstein-Langevin equation, will be relevant if the constant, it saturates to, is comparable to the expectation values appearing in the semiclassical analysis. Thus, in principle, these space-times are vulnerable to long range effects. 

This is a bit interesting as, in the late time limit, the Wightman function (and hence the stress energy correlator) drops the time (or the scale factor) dependency. 
For constant time-sheets, we have 
\begin{equation}
G^{P.L.}(\eta,\vec{x},\eta',\vec{x}') = \frac{H^2(H\eta)^{2q-2}}{16\pi^2} _2F_1(2+q,1-q,2,1-\frac{(\Delta \vec{x})^2}{4\eta^2}) \, .
\end{equation}
In the $\eta \to 0$ limit, we have 
\begin{multline}
G^{P.L.}(\eta,\vec{x},\eta',\vec{x}') = \frac{H^2(H\eta)^{2q-2}}{16\pi^2} \Gamma(2+q)\Gamma(1-q)\\ \Big[\frac{\Gamma(-1-2q)(\big(\frac{(\Delta \vec{x})^2}{4\eta^2}\big))^{-2-q}}{\Gamma(1-q)\Gamma(-q)} \sum_{k =0}^{\infty}\frac{(2+q)_k(1+q)_k\big(-\frac{(\Delta \vec{x})^2}{4\eta^2}\big)^{-k}}{k! (2+2q)_k} \\
+ \frac{\Gamma(1+2q)\big(\frac{(\Delta \vec{x})^2}{4\eta^2}\big)^{-1+q}}{\Gamma(2+q)\Gamma(1+q)}\sum_{k =0}^{\infty}\frac{(1-q)_k(-q)_k\big(-\frac{(\Delta \vec{x})^2}{4\eta^2}\big)^{-k}}{k! (-2q)_k}\Big].
\end{multline}
Since $a(\eta)= (H\eta)^{-q}$ (i.e., $H\eta = a^{-1/q}$), we can convert the above expression in terms of the physical distance on constant time sheets, i.e. $a^2(\Delta \vec{x})^2$, and in terms of $a(\eta)$ i.e., 
\begin{multline}
G^{P.L.}(\eta,\vec{x},\eta',\vec{x}') = \frac{H^2}{16\pi^2} \Gamma(2+q)\Gamma(1-q)\\ \Big[\frac{\Gamma(-1-2q)\big(\frac{H^2}{4}\big)^{-2-q} a^{2q-2/q}}{\Gamma(1-q)\Gamma(-q)(a^2(\Delta \vec{x})^2)^{2+q}} \sum_{k =0}^{\infty}\frac{(2+q)_k(1+q)_k\big(-\frac{H^2}{4}\big)^{-k}(a^2(\Delta \vec{x})^2)^{-k}(a^{-2+2/q})^{-k}}{k! (2+2q)_k}\\
+ \frac{\Gamma(1+2q)\big(\frac{H^2}{4}\big)^{-1 +q}a^{2-2q}}{\Gamma(2+q)\Gamma(1+q)(a^2(\Delta \vec{x})^2)^{1-q}}\sum_{k =0}^{\infty}\frac{(1-q)_k(-q)_k\big(-\frac{H^2}{4}\big)^{-k}(a^2(\Delta \vec{x})^2)^{-k}(a^{-2+2/q})^{-k}}{k! (-2q)_k}\Big] .
\end{multline}

One can check that the leading term of the second series in the square bracket is the dominant term for $q > -1/2$, in the $\eta \to 0$ limit, which kills off all $a$ dependence at late times, assuming a pseudo-conformal form. It is worth noting that, for all prior times, there  is a $\eta-$ dependency in the expression, which gradually decays and at the end we are left with the constant leading order term. Therefore, long distance correlators, with small co-ordinate values, of this space-time maintain the initial time correlations.  

\item $ q = 0$ :  This is a special limit of no dynamics i.e., $a(\eta) =1$, and hence is the flat space result, which is well studied \cite{Padmanabhan:1988cj, Ford:2005sp, Frob:2013sxa}. 
The Wightman function for Minkowskian space-time is given by $G(x,x') = \frac{1}{4\pi^2 (-(\eta-\eta')^2 + (\Delta\vec{x})^2)}$. Using this expression, we find that the noise kernel, on constant time-sheets for finite spatial distance, is given by:
\begin{equation}
\langle \hat{t}_{00}(\eta,\vec{x})\hat{t}_{00}(\eta,\vec{x'}) \rangle_{P.L.} = \frac{3}{2\pi^4(\Delta \vec{x})^8}  \, .
\end{equation}
Evidently, for constant co-moving distance, the correlator survives as the co-moving and physical distances are the same and physical distance does not grow in ``late time'' or ``early time'' because of lack of dynamics. For large physical distance, there is no appreciable stochastic effect.

\item  $ q \in (-2,0)$ : 
In this case, $a(\eta) = (H\eta)^{|q|}$ and hence the late time universe corresponds to  $\eta \to \infty$. For this case, we have 
\begin{multline}
\langle t_{00}(\eta,\vec{x})t_{00}(\eta,\vec{x'}) \rangle_{P.L.} \big|_{\text{late time}}= \lim_{\eta \to \infty} (H\eta)^{4q-4}\Bigg[\frac{3H^4\eta^{4}}{2\pi^4(\Delta \vec{x})^8} + \frac{\eta^2H^4(3q + 4q^2)}{8\pi^4(\Delta \vec{x})^6} \\ \\ + \frac{H^4 q}{64\pi^4(\Delta \vec{x})^4}\Big((-4 - 7q + 6q^2 + 11q^3) \\ \\  + 2(1+q)(-1+q)^2\big[ 2\gamma  + log\big(\frac{(\Delta \vec{x})^2}{4\eta^2}\big) + \psi^{(0)}(1-q) + \psi^{(0)}(2+q))\big]\Big) +  O(\eta^{-2})\Bigg] .
\end{multline}
Here $\gamma$ is Euler gamma symbol and $\psi^{(0)}(z)$ is PolyGamma function. Clearly, the late time correlator has a behaviour ${\cal O}(\eta^{4 q})$ for fixed $\Delta x$ which washes away any quantum correlation at late times.

 \item $q = -2$ :  This case is particularly interesting as we see that, $m^2 = H^2(1-q)(2+q)|\rightarrow 0$ for $q\rightarrow -2$, and hence, a massless scalar field in this particular Friedmann space-time is conformally equivalent to a massless scalar field in de Sitter space-time. The stress energy correlator for this case is given as
\begin{multline}
\langle t_{00}(\eta,\vec{x})t_{00}(\eta,\vec{x'}) \rangle_{P.L.} \big|_{\text{late time}} = \lim_{\eta \to \infty} \lim_{\epsilon \to 0}H^{-12}\Bigg[\frac{3H^4}{2\pi^4\eta^8 (\Delta \vec{x})^8} + \frac{4}{(\Delta \vec{x})^6 \eta^{10}}\Big(\frac{5H^4}{16\pi^4} + O(\epsilon)\Big) \\ \\ + \frac{1}{\eta^{12} (\Delta \vec{x})^4}\Big(\frac{9H^4}{16\pi^4 \epsilon} + \frac{9(6 H^4 +  H^4log(\frac{(\Delta \vec{x})^2}{4\eta^2}))}{16\pi^4}+ O(\epsilon)\Big)  \\ \\ + \frac{1}{4(\Delta \vec{x})^2\eta^{14}}\Big(-\frac{27H^4}{8\pi^4\epsilon} - \frac{27(7H^4 + 2H^4 log(\frac{(\Delta \vec{x})^2}{4\eta^2}))}{16\pi^4} + O(\epsilon)\Big) \\ \\ + \frac{1}{16\eta^{16}}\Big(\frac{81H^4}{8\pi^4\epsilon^2} + \frac{27H^4(10 + 3  log(\frac{(\Delta \vec{x})^2}{4\eta^2})))}{4\pi^4\epsilon} + O(\epsilon^0)\Big) + O(\eta^{-18})\Bigg].\\
\end{multline}
Since $\epsilon \to 0$ limit blows up for all large but finite $\eta$, the long range correlators become dominant over the expectation values and one needs to resort to stochastic gravity necessarily. In fact, it is easy to show that such divergent behaviour persists at all times.
This is not unexpected as we have already seen that the Wightman function diverges secularly for massless case in de Sitter. However, $q=-2$ space-time is connected to the de Sitter case as  
\begin{equation}
G^{q=-2}_{m=0}(x,x')=(H^2\eta \eta')^{-3}G^{dS}_{m=0}(x,x'),
\end{equation}
and thus, in this space-time, the divergent term from the de Sitter, develops time dependence and survives under derivative actions in Eq.(\ref{correlator}). Similar space-time dependent divergence appears for universes with $q <-2$ and $q>1$ corresponding to $|\nu|>3/2$. Therefore, the semiclassical (or even classical) analysis on these universes is potentially unstable in the face of quantum fluctuations.
\end{itemize}

\section{Energy-Energy Correlation on constant time-sheets}\label{EnergyCorr}
In the previous section, we evaluated the $\langle t_{0000}\rangle$ component of the noise kernel for different Friedmann space-times. We realize that, in some cases, the late time character cares only for the co-ordinate separation ($\Delta x$) which is not coordinate invariant. This is not  unexpected as the noise kernel is not an invariant scalar. However, one can construct invariant scalars out of these to assess the effect of stochastic fluctuations more covariantly. For this purpose, we consider the behaviour of energy-energy density correlator in the late-time universe in these Friedmann universes. Energy density at any point, $x$, is given by $T_{\alpha\beta}(x)t^{\alpha}t^{\beta}$, where $t^{\alpha}$ is some time-like vector. So, if we consider a co-moving time-like path $x(\lambda) = (N(\lambda), \vec{x} = \vec{c})$, then we see that $t^{\alpha} = (\dot{N}(\lambda), \vec{0})$ and hence the unit parametrization in de Sitter space implies that $\dot{N}(\lambda) = 1/a(\eta)$. This implies that the energy density, at point $(\eta,\vec{x})$, is $T_{00}(\eta,\vec{x})/(a{(\eta)})^2$ and the energy-energy density correlator between the points $(\eta,\vec{x})$ and $(\eta',\vec{x}')$ is given by:
\begin{equation}\label{eq11}
\frac{\langle t_{00}(\eta,\vec{x})t_{00}(\eta',\vec{x}')\rangle}{(a(\eta)a(\eta'))^2} \, .
\end{equation}
Using the expressions for $\langle t_{00}(\eta,\vec{x})t_{00}(\eta',\vec{x}')\rangle$ obtained  in the previous section, we can obtain the energy density correlators, over large physical distances, for different Friedmann universes.
\begin{itemize}
\item $q=1$  : This is the de Sitter space-time. Using equation (\ref{eq11}), we see that the energy-energy correlator, in the $\eta\to 0$ limit, is $0$ for every value of $\nu$ except for $\nu = 3/2$, for which it is constant ($= 9H^8/128\pi^4$). One point to note is that, for $\nu = 3/2$, the infra-red problem doesn't appear at the level of noise kernel, as we take the massless field as the klimiting case. Since the massless fields have no de Sitter invariant vacuum \cite{Allen:1985ux}, one needs to regularize the divergent piece in the Wightman function in a proper way. However, it can be shown that the regularized massless Wightman function does not yield any non-zero energy density correlator in late time limit \cite{Padmanabhan:1988se}.

\item $ q \in (0,1)$ : 
Using Equation $(\ref{eq11})$ for the energy-energy correlator and the fact that $a(\eta) = (H\eta)^{-q}$, we see that, in this case, the energy-energy correlator goes to $0$ in the $\eta \to 0$ limit.

\item $ q = 0$ : This is the flat space case i.e., $a(\eta)=1$, where the $\langle t_{0000}\rangle$ is same as the energy-energy correlator. 

\item  $ q \in (-2,0)$ : 
In this case, we find that the late time universe i.e., $\eta \to \infty$, has zero value for the energy-energy correlator.

\item  $q = -2$ : 
Again using equation $(\ref{eq11})$, we see that the energy-energy correlator, for this case,  is divergent in the  late time limit (in fact, for arbitrary values of $\eta$) as there is a pole in the Wightman function at $\nu=3/2$.  Similarly the energy energy correlator will blow up for $q<-2$  and $q>1$ too.
\end{itemize}
Thus, we see the coordinate invariant quantity is mostly under control and may give rise to finite value observable quantities. However, some Friedmann universes still give divergent energy-energy correlators which are explained below. 

\subsection{Cosmological Implications:}
The equation of state parameter,$\displaystyle w$, for an ideal fluid driving the Friedmann universe, is related to the exponent of the scale factor in Friedmann cosmologies by $q = -2/(1+3\displaystyle w)$ (see \cite{Lochan:2018pzs}).
Different values of $\displaystyle w$ represent the dominance of different types of fluid during the evolution of the universe e.g., $\displaystyle w =0$ corresponds to dust whereas $\displaystyle w = 1/3$ corresponds to radiation. Therefore, we see that the Friedmann universe with $q \in (0,1]$ is driven by a fluid with equation of state parameter $\displaystyle w \in (-\infty, -1]$  whereas $\displaystyle w \in [0,\infty)$ for $q \in [-2,0)$. This means that the quantum fluctuations ($(a=0,b=0,c=0,d=0)$ noise kernel component) may remain relevant for  $\displaystyle w \in (-\infty, -1]$ in the late-time universe where the initial quantum fluctuation remain frozen under expansion,  whereas there are no quantum fluctuations left over for $\displaystyle w \in (0,\infty)$ in the late-time universe.  Interestingly, for $q=-2$ i.e., $\displaystyle w =0$, a pressureless dust driven universe is the one, most affected by stochastic fluctuations as the noise kernel components (or even scalar correlators) blow up for this space-time. Thus any semi-classical or possibly even classical analysis in dust driven universe is subject to scrutiny under Stochastic correction. In other words, this space-time remains as quantum as ever. The same remains true for $q < -2$ or $q > 1$ corresponding to $\displaystyle w \in [-1,0]$. Thus, the space-times in these regime never drop their quantum character and a higher order quantum analysis is necessary. Interestingly, accelerating universes require $\displaystyle w < -1/3$ and hence, any accelerating universe also seeks for a quantum treatment. The summary of this section is given by the following diagram:
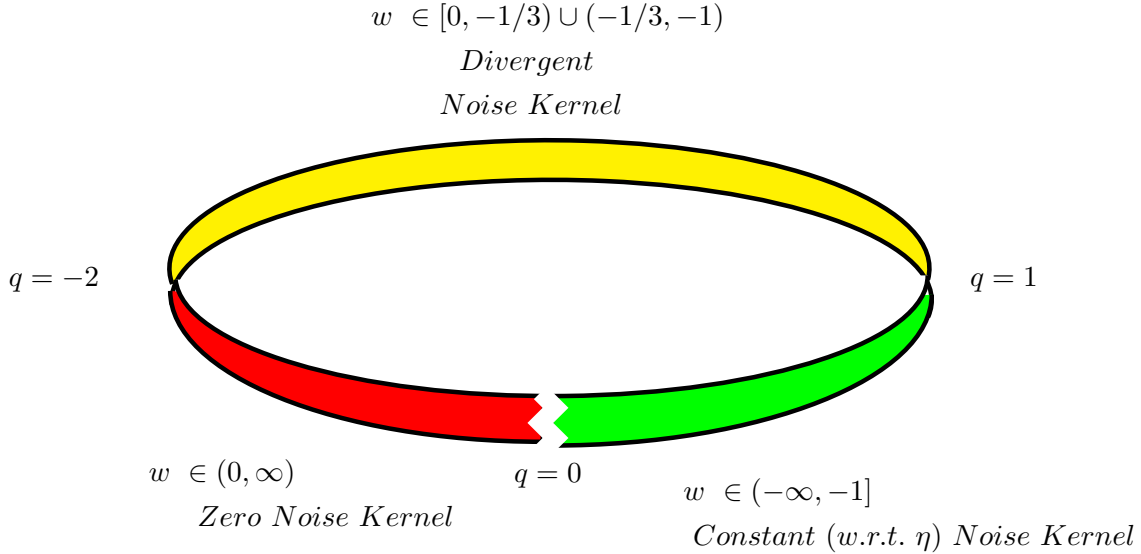
\begin{figure}[h]
\begin{tikzpicture}[line width=0.6mm]
\begin{scope}
 \draw (0,4.5) node {$\begin{array}{{>{\displaystyle}l}}
\displaystyle w \ \in [0,-1/3)\cup (-1/3, -1)\\
\ \ \ \ \ \ \ \ \ Divergent\ \\
\ \ \ \ \ \ \ Noise\ Kernel\ 
\end{array}$}       ;
\draw (-6.5,1.6) node{$q = -2$};
\draw (6,1.6) node{$q = 1$};
\draw (0,-1) node{$q = 0$};
\draw (-3.5,-1.25) node{$\displaystyle  \begin{array}{{>{\displaystyle}l}}
\ \ \ \ \displaystyle w \ \in ( 0,\infty )\\
\ \ \ \ \ \ \ \ \ Zero\ Noise\ Kernel\ 
\end{array}$};
\draw (4.5,-1.5) node{$\displaystyle  \begin{array}{{>{\displaystyle}l}}
\ \ \ \ \displaystyle w \ \in ( -\infty ,-1]\\
\ \ \ \ \ Constant\ (w.r.t.\ \eta)\ Noise\ Kernel\  
\end{array}$};
      \path [draw=black,fill = yellow] (5,1.6) arc (-5:187: 5cm and 1.7cm);
      \path [draw=black,fill = white] (5,1.1) arc (-8:182:5cm and 1.6cm);
      \path [draw=black,fill = red] (-4.97,1.45) arc (180:270: 4.9cm and 2.0cm)--(0.0,1.1);
      
      \path [draw=black,fill = green] (5.05,1.40) arc (0:-90:5cm and 2 cm)--(0.0,1.1);
      \path [draw=black,fill = white] (-4.9,1.6) arc (182:360: 4.95 cm and 1.6cm);
      \draw[white,line width=2.4 mm] (-0.1,0.1) -- (0.1,-0.1)--(-0.1,-0.3)--(0.1,-0.5)--(-0.1,-0.7);

    \end{scope}
\end{tikzpicture}
\caption{Relation between different types of fluid (and the corresponding Friedmann space-times) and the behaviour of noise kernel in these regions.}
\end{figure}

\section{Conclusions}\label{Summary}
In this paper, we analyse the stability of various Friedmann universes under the possible effect of stochastic correction term in the Einstein-Langevin equation. Using the relations between Wightman functions in de Sitter and Friedmann universes, as well as the relation of the Wightman function to noise kernel components, we relate the noise kernel in Friedmann universe to the conformal scaling of the noise kernels of massive fields in de Sitter universe.  Typically quantum fluctuations are expected to decay over large length scales in flat space-time and remain relevant only over extremely small scales. A Friedmann universe is conformally flat, i.e. the points which are initially very close by, will get physically separated under a global topological expansion. However, in this lies an interesting possibility, where two space-time points are physically apart by large distances while maintaining small co-ordinate separation. Under certain scenarios, e.g. for conformal fields, it may be possible that the noise kernel cares about the co-ordinate separation and not about the true physical distances. In those cases, the quantum fluctuations which were stronger when the points had not accelerated away from each other, remain as strong under time (and hence physical distance) growth.  Further, there can be cases, where the signature of small co-ordinate distance get enhanced with increasing conformal scale factor. We argue that certain Friedmann universes develop this tendency in the late time era, and thus maintain potentially significant second order correction to the semiclassical equations. We first list our findings in this paper as follows:
\begin{itemize}
\item \textbf{Minimally Coupled Massive Scalar Field in de Sitter Space-time :} We first consider a minimally coupled massive scalar field on $dS_4$ and study the variation of the $(a=0, b=0, c=0, d=0)$ component of the noise kernel as a function of mass. We consider only those cases in which $|\nu|$ lies in the range $[0,3/2]$ because this range of values for $\nu$ ensures that the mass is real as $m^2/H^2 = 9/4 - \nu^2$. If we consider finitely separated points on constant time sheets and take the late time, $\eta \to 0$, limit (for which the scale factor, $1/(H\eta)$, grows), we find that the considered component of noise kernel undergoes a sort of phase  transition from zero to non-zero values, with the critical value being $\nu  = 1/2$. The noise kernel components stay vanishing for $\nu \in [0,1/2)$ and assume finite non-zero value at $\nu = 1/2$. Further, for $\nu >1/2$, it diverges in the limit $\eta \to 0$. We also show that the energy-energy correlator (which is just the $(a =0,b=0,c=0,d=0)$ component of the noise kernel suppressed by the fourth power of the scale factor) vanishes for all values of $\nu$ except for $\nu = 3/2$ for which it is finite (at the value $9H^8/128\pi^4$).

\item \textbf{Non-Minimally Coupled Scalar Field in de Sitter Space-time:} We also evaluate the stochastic corrections for the  non-minimally coupled massive scalar field  on $dS_4$, which amounts to adding $\xi R \phi^2$ to the minimally coupled Lagrangian. In this case as well, we consider finitely separated points on constant time sheets in the $\eta \to 0$ limit. We show that the variation of the $(a=0, b=0, c=0, d=0)$ component of the noise kernel as a function of $\nu$ is similar to the minimally coupled  case with $\nu \rightarrow \nu  = \sqrt{9/4 - (m^2 + 12\xi H^2)/H^2}$. In effect, what we have shown is that the considered component is 0 for $(m^2/H^2 + 12 \xi)>2$ and becomes non-zero  for $(m^2/H^2 + 12 \xi)=2$ and diverges for $(m^2/H^2 + 12 \xi)<2$. Therefore, conformal coupling $\xi =1/6$ cures the divergences for all masses. Similarly,  the energy-energy correlator stays 0 for all values of  $m^2/H^2 + 12\xi$ in the range $[0,\frac{9}{4})$ but becomes constant  for the $m^2/H^2 + 12\xi = 0$. 

\item \textbf{Conformally related massless Scalar Field in Friedmann Space-times:} Using the results of \cite{Lochan:2018pzs}, that establish an equivalence between a massless scalar field in Friedmann space-times with that of a massive scalar field in de Sitter space where the mass gets related to the exponent of the scale factor of Friedmann, we compute the noise kernels for various Friedmann universes. For $\nu$ lying between $[-3/2, 3/2]$, $q$ can take values between $[-2,1]$. Since we are interested in large physical separation (late time universes), the late time limit corresponds to $\eta \to 0$ for $q \in (0,1]$ whereas it corresponds to $\eta \to \infty$ for $q \in [-2,0)$. We find that, for $q \in [0,1)$, the considered noise kernel component approaches a constant value while, for $q \in (-2,0)$, it vanishes over large physical separations. Similarly, for the energy-energy correlator, we showed that it is zero for $q \in (-2,0)\cup (0,1)$.  However,  particularly interesting cases are for $q \leq -2$ and $q\geq 1$(which correspond to universes driven by $-1 <\displaystyle w < 0$ equation of state fluids) where, as shown in \cite{Lochan:2018pzs}, the Wightman function has a divergent term with space-time dependence. The conformal connection between the fields of de Sitter and Friedmann spaces provides an additional conformal time dependence to the divergent term in the de Sitter Wightman function, which contributes dominantly in both the noise kernel and the energy-energy correlators. Therefore, the universes, which are driven by such fluids, remain susceptible to quantum fluctuations at late time as well.
\end{itemize}

Several observations and their  implications are in order. First we see that the space-times which are potentially stable against the stochastic corrections are those driven by $\displaystyle w>0$ or phantom universes $\displaystyle w<-1$ if expectation values are large. This reinforces the quantum fluctuation structure suggested in 
\cite{Lochan:2018pzs}. However, there are a couple of more interesting points to be learnt from this exercise. Firstly, in the de Sitter case, the late time divergence for fields with $\nu >1/2$ is dynamic in nature (unlike the Wightman function which is space-time independent, as well as, only for massless case). Secondly, in the phantom space-times, the noise kernel does not grow out of control but still maintains a non vanishing noise over large length scales if the points were born close by. Further, for $q = -2,$ we see that there is a non-dynamic divergence in all correlators, which makes dust driven universe potentially unstable under quantum noise.  From power counting arguments, it is easy to visualize that, for space-times which are driven by $\displaystyle w \in (-1,0 ]$, the quantum fluctuations will always remain important, in a similar spirit. This has many physical implications on semiclassical physics in such space-times, e.g., quasi-de Sitter inflation or late time quintessence field driven universes. Therefore, the dynamics or the growth of perturbations of the massless kind in these space-times needs to be evaluated carefully, accounting for such corrections. The effects of stochastic corrections on background as well as perturbation dynamics in these space-times are currently being studied and will be reported elsewhere. Besides this, cosmological data suggests that $\displaystyle w$ is a dynamical quantity i.e., it changes with time. In this paper, we have studied these stochastic corrections only for constant $\displaystyle w$ driven universes and as such our analysis requires some changes to take this fact into account which we plan to study in future. However, if the time variation of $\omega$ is sufficiently small compared to the rate of expansion of the universe i.e., $\dot{\omega}/\omega << \dot{a}/a$, then our study becomes relevant in the dust driven as well as dark energy dominated  or quintessence era etc.

\section*{Acknowledgments}
AD would like to acknowledge the financial support from University Grants Commission, Government of India, in the form of Junior Research Fellowship (UGC-CSIR JRF/Dec-2016/510944). Research of KL is partially supported by the Department of Science and Technology (DST) of the Government of India through a research grant under INSPIRE Faculty Award (DST/INSPIRE/04/2016/000571). The authors thank T. Padmanabhan for providing useful comments on the manuscript.
 
\appendix 
\section{Basic Computation on constant time surface}\label{CorrDerr}
In this appendix, we collect certain results which are important in some of the calculations presented in the main text of this paper. For de-Sitter invariant vacuum, $G(x,x') = G(Z(x,x'))$, we see, using equation (\ref{eq6}), that
\begin{eqnarray} 
\nabla_{\mu}'G & = &  G' \Big[\frac{(x-x')_{\mu}}{\eta\eta'} + \frac{\Delta s^2}{2\eta\eta'^2}\delta_{\mu 0} \Big]\, , \\
\nabla_{\nu}G & = & G' \Big[-\frac{(x-x')_{\nu}}{\eta\eta'} + \frac{\Delta s^2}{2\eta^2\eta'}\delta_{\nu 0} \Big] \, , \\
\nabla_{\nu}\nabla_{\mu}'G & = & G''\Big[\frac{(x-x')_{\mu}}{\eta\eta'} + \frac{\Delta s^2}{2\eta\eta'^2}\delta_{\mu 0} \Big]\Big[-\frac{(x-x')_{\nu}}{\eta\eta'} + \frac{\Delta s^2}{2\eta^2\eta'}\delta_{\nu 0} \Big] \nonumber \\ & + & G'\Big[\frac{\eta_{\mu \nu}}{\eta\eta'}-\frac{(x-x')_{\mu}}{\eta^2\eta'}\delta_{\nu 0} + \frac{(x-x')_{\nu}}{\eta\eta'^2}\delta_{\mu 0} - \frac{\Delta s^2}{2\eta^2\eta'^2}\delta_{\nu 0}\delta_{\mu 0 } \Big]  \, .
\end{eqnarray} 
On constant time sheets i.e., $\eta = \eta'$, we have
\begin{eqnarray}
 \,  Z(x,x') & = &  1 - \frac{(\Delta \vec{x})^2}{2\eta^2} , \, \\
 \nabla_{i}'G &=& G'\Big[\frac{(x - x')_{i}}{\eta^2}\Big] \, , \\
 \nabla_{i}G &=& G'\Big[-\frac{(x - x')_{i}}{\eta^2}\Big] \, ,  \\
\nabla_{0}'G & = & G'\Big[\frac{(\vec{x} - \vec{x}')^2}{2\eta^3}\Big]  \, , \\
\nabla_{0}G &=&  G'\Big[\frac{(\vec{x} - \vec{x}')^2}{2\eta^3}\Big] \, , \\
\nabla_{i}\nabla_{j}'G &=& G'' \Big[-\frac{(x-x')_i(x-x')_j}{\eta^4}\Big] + G'\Big[\frac{\delta_{ij}}{\eta^2}\Big] \, ,  \\
\nabla_{0}\nabla_{j}'G &=& G'' \Big[\frac{(x-x')_j(\vec{x}-\vec{x}')^2}{2\eta^5}\Big] - G'\Big[\frac{(x-x')_j}{\eta^3}\Big]  \, , \\
\nabla_{i}\nabla_{0}'G &=& G'' \Big[-\frac{(x-x')_i(\vec{x}-\vec{x}')^2}{2\eta^5}\Big] + G'\Big[\frac{(x-x')_i}{\eta^3}\Big]   \, , \\
 \nabla_{0}\nabla_{0}'G &=&G'' \Big[\frac{(\vec{x}-\vec{x}')^4}{4\eta^6}\Big]  +  G'\Big[-\frac{1}{\eta^2} - \frac{(\vec{x}-\vec{x}')^2}{2\eta^4}\Big] 
  \, .
\end{eqnarray}

\section{ Power Counting for Noise Kernel:}\label{PowerCount}
\subsection{Minimal Coupling}
In this appendix, we present a power counting argument to find out for what values of $\nu$ the noise kernel for de Sitter space-time i.e., the equation(\ref{eq7}), diverges as $\eta \to 0$ (late time universe). If we look at the first term in the equation(\ref{eq7}) i.e., 
\begin{equation}
(G'')^2\Big[\frac{(\Delta \vec{x})^6}{4 \eta^{10}} + \frac{(\Delta \vec{x})^8}{32\eta^{12}} + \frac{(\Delta \vec{x})^4}{2\eta^{8}}\Big] \, ,
\end{equation}
we see that the most divergent term in the square brackets is $O(\eta^{-12})$. So, if we can find the values of $\nu$ for which the least power of $\eta$ in $(G'')^2$ is $<12$, then we have found the range of $\nu$ for which this term diverges. \\
Since the Wightman function and its derivatives are functions of $\frac{1+Z}{2}\big( = 1 -\frac{(\Delta \vec{x})^2}{4\eta^2}\big)$ \footnote{See equations(\ref{eq8}), (\ref{eq9}), (\ref{eq10})}, we must look at the series expansion of the Wightman function and its derivative at large values of their arguments in the $\eta \to 0$ limit. If we look at the following series expansion of $_2F_1(a,b,c,z)$ \cite{abramowitz+stegun} (valid for large $|z|$ and $a-b \notin \mathbb{Z}$)\footnote{In our case, $a -b = 2\nu$ which is not an integer for every value of $\nu$ in the range $[0,\frac{3}{2}]$ except for $\nu = 0,\frac{1}{2},1,\frac{3}{2}$. But we have already considered these cases separately in the section 3.}:
\bea
_2F_1(a,b,c,z) = \frac{\Gamma(b-a)\Gamma(c)(-z)^{-a}}{\Gamma(b)\Gamma(c-a)}\sum_{k =0}^{\infty}\frac{a_k(a-c+1)_kz^{-k}}{k! (a-b+1)_k} \nonumber\\
+ \frac{\Gamma(a-b)\Gamma(c)(-z)^{-b}}{\Gamma(a)\Gamma(c-b)}\sum_{k =0}^{\infty}\frac{b_k(b-c+1)_kz^{-k}}{k! (-a+b+1)_k} \, ,
\eea
and keep in mind equation (\ref{eq10}), we find that the least power of $\eta$ in $(G'')^2$ is $14 -4 \nu$. Therefore, the above term diverges for $\nu > \frac{1}{2}$. A similar analysis with the other terms in the equation (\ref{eq7}) tells us that the equation (\ref{eq7}) diverges for $\nu >\frac{1}{2}$.\\
These arguments can be applied to the general components of the noise kernel. In fact, looking at the least powers of $\eta$ in the  formulae listed in Appendix A for different covariant derivatives of Wightman function on constant time sheets and the equation (\ref{eq5}), we see that 

\begin{table}[h]
\begin{tabular}{|c|c|}
\hline
$\langle t_{ab}(\eta,\vec{x})t_{cd}(\eta,\vec{x'})>_{dS}$ & \begin{tabular}[c]{@{}c@{}}Leading order \\    behaviour in \\      $\eta$\end{tabular} \\ \hline
$a = 0, b = 0, c= 0, d =0 $                        &     $O(\eta^{2-4\nu})$                                                                                    \\ \hline
$a = 0, b = 0, c= 0, d =l $                        &     $O(\eta^{3-4\nu})$                                                                                    \\ \hline
$a = 0, b = j, c= 0, d =0 $                        &     $O(\eta^{3-4\nu})$                                                                                    \\ \hline
$a = 0, b = j, c= 0, d =l $                        &     $O(\eta^{4-4\nu})$                                                                                    \\ \hline
$a = i, b = j, c= 0, d =0 $                        &     $O(\eta^{2-4\nu})$                                                                                    \\ \hline
$a = 0, b = 0, c= k, d =l $                        &     $O(\eta^{2-4\nu})$                                                                                    \\ \hline
$a = i, b = j, c= k, d =0 $                        &     $O(\eta^{3-4\nu})$                                                                                    \\ \hline
 $a = 0, b = j, c= k, d =l $                        &     $O(\eta^{3-4\nu})$                                                                                    \\ \hline     
 $a = i, b = i, c= k, d =l \ \ and \ \ k \neq l$                        &     $O(\eta^{4-4\nu})$                                                                                    \\ \hline
  $a = i, b = j, c= k, d =k \ \  and \ \ i \neq j$                        &     $O(\eta^{4-4\nu})$                                                                                    \\ \hline
   $a = i, b = i, c= k, d =k $                        &     $O(\eta^{2-4\nu})$                                                                                    \\ \hline
    $a = i, b = j, c= k, d =l \ \ and \ \ i \neq j, k \neq l $                        &     $O(\eta^{6-4\nu})$                                                                                    \\ \hline
\end{tabular}
\end{table}
\newpage
\subsection{Non-minimal coupling}\label{NonMinimal}
Below is given the expression of the noise kernel for the non-minimally coupled massive scalar field on de Sitter space-time in terms of the Wightman function and its covariant derivatives. First we substituted the expression (\ref{eq16}) for stress-energy tensor in the definition of the noise kernel (\ref{eq15}). Since the definition of the noise kernel contains the vacuum expectation of product of two stress-energy tensors and each stress-energy operator contains two field operators, we would get the vacuum expectation of product of four field operators. Then, we can use the Wick theorem to express this vacuum expectation as the product of two Wightman function and obtains the equation (\ref{eq14}). We obtain the below given expression by substituting the expressions (\ref{eq12}), (\ref{eq13}) for $P_{ab}(x,y)$ and $M_{ab}(x,y)$ in equation(\ref{eq14}) 
\begin{multline}
\langle t^{nm}_{ab}(x)t^{nm}_{cd}(x')\rangle = \Bigg[(1-2\xi)^2\big(\nabla_{b}\nabla_{c}'G(x,x')\nabla_{a}\nabla_{d}'G(x,x') + \nabla_{b}\nabla_{d}'G(x,x')\nabla_{a} \nabla_{c}'G(x,x')\big) \\- (1 - 4\xi)(1 -2\xi)\eta_{cd}\eta^{\rho \sigma}\nabla_{a}\nabla_{\rho}'G(x,x')\nabla_{b}\nabla_{\sigma}'G(x,x') - \frac{(m^2 + 6H^2\xi)(1-2\xi)}{H^2\eta'^2} \eta_{cd}\nabla_{a}G(x,x')\nabla_{b}G(x,x') \\ -(1 - 4\xi)(1 -2\xi)\eta_{ab}\eta^{\gamma \delta}\nabla_{\gamma}\nabla_{c}'G(x,x')\nabla_{\delta}\nabla_{d}'G(x,x') +  \frac{(1-4\xi)^2}{2}\eta_{ab}\eta^{\gamma \delta}\eta_{cd}\eta^{\rho \sigma}\nabla_{\gamma}\nabla_{\rho}'G(x,x')\nabla_{\delta}\nabla_{\sigma}'G(x,x') \\ 
 + \frac{(m^2 + 6H^2\xi)(1-4\xi)}{2H^2\eta'^2}\eta_{ab}\eta^{\gamma \delta}\eta_{cd}\nabla_{\gamma}G(x,x')\nabla_{\delta}G(x,x') - \frac{(m^2 + 6H^2\xi)(1-2\xi)}{H^2\eta^2} \eta_{ab}\nabla_{c}'G(x,x')\nabla_{d}'G(x,x') \\ + \frac{(m^2 + 6H^2\xi)(1-4\xi)}{2H^2\eta^2} \eta_{ab}\eta_{cd}\eta^{\rho \sigma}\nabla_{\rho}'G(x,x')\nabla_{\sigma}'G(x,x') + \frac{1}{2H^4\eta^2\eta'^2}(6H^2 \xi + m^2)^2 \eta_{ab}\eta_{cd}G^2\Bigg] \\ +  2\xi\Bigg[2\eta_{cd}(1-2\xi)\big(\nabla_{(a}G\nabla_{b)}\Box^{'}G\big) - 2(1-2\xi)\big(\nabla_{(a}G\nabla_{b)}\nabla_{(c}'\nabla_{d)}'G\big) - \frac{(6H^2\xi + m^2)}{(H\eta)^2}\eta_{ab}\eta_{cd}G\Box'G \\ 
-(1-4\xi)\eta_{ab}\eta_{cd}(\eta^{rs}\nabla_{s}G\nabla_{r}\Box'G) + (1-4\xi)\eta_{ab}(\eta^{rs}\nabla_{s}G\nabla_{r}\nabla_{(c}'\nabla_{d)}'G)  + \frac{(6H^2\xi + m^2)}{(H\eta)^2}\eta_{ab}G\nabla_{(c}'\nabla_{d)}'G\Bigg] \\ +  2\xi\Bigg[2\eta_{ab}(1-2\xi)\big(\Box\nabla_{(c}'G\nabla_{d)}'G\big) - 2(1-2\xi)\big(\nabla_{(a}\nabla_{b)}\nabla_{(c}'G\nabla_{d)}'G\big) - \frac{(6H^2\xi + m^2)}{(H\eta')^2}\eta_{ab}\eta_{cd}G\Box G \\ 
-(1-4\xi)\eta_{ab}\eta_{cd}(\eta^{rs}\Box\nabla_{s}'G\nabla_{r}'G) + (1-4\xi)\eta_{cd}(\eta^{mn}\nabla_{(a}\nabla_{b)}\nabla_{n}'G\nabla_{m}'G)  + \frac{(6H^2\xi + m^2)}{(H\eta')^2}\eta_{cd}G\nabla_{(a}\nabla_{b)}G\Bigg] \\ + 4\xi^2 \Bigg[\eta_{ab}\eta_{cd}\big(\Box 'G\Box G +  G \Box \Box' G \big) - \eta_{ab}\big(\nabla_{(c}'\nabla_{d)}'G \Box G + G\Box \nabla_{(c}'\nabla_{d)}'G\big) \\ - \eta_{cd}\big(\nabla_{(a}\nabla_{b)}G \Box' G + G\nabla_{(a}\nabla_{b)} \Box' G\big)  + \big( \nabla_{(a}\nabla_{b)}G\nabla_{(c}'\nabla_{d)}'G + G\nabla_{(a}\nabla_{b)}\nabla_{(c}'\nabla_{d)}'G\big)\Bigg]  \, .
\end{multline}
The first square bracket contains the $P_{ab}P_{cd}$ term, the second and the third square brackets contain the $P_{ab}M_{cd}$ and $M_{ab}P_{cd}$ terms respectively. Whereas the fourth square bracket contains the $M_{ab}M_{cd}$ term. We can use the same power counting analysis as is done for the minimal coupling section of this appendix and study the behaviour of divergences for noise kernel as function of mass and the coupling constant $\xi$.

\section{Divergence in noise kernel for $\omega \in (-1,0)$ driven universe}\label{NegativeEoS}
Looking at the equation $(85)$ of \cite{Lochan:2018pzs}, we see that the Wightman function for massless scalar field in Friedmann space-times is given by : 
\begin{equation}
G(x,x')= \frac{\beta^2 (\eta\eta')^{q-1}}{8\pi^2}\int_{0}^{\infty}ds\frac{ s^{\frac{1}{2} -\nu}}{(s^2 - 2 Z s + 1)^{\frac{3}{2}}}
\end{equation} 
\subsection*{Case $\nu < -\frac{3}{2}$:}
Now, consider the integral for large $s$ values i.e., 
\begin{multline}
G(x,x')= \frac{\beta^2 (\eta\eta')^{q-1}}{8\pi^2}\Big[finite\ term + \int_{N}^{\infty}ds\ s^{-\frac{5}{2} -\nu}\Big(1 - 2\frac{Z}{s} + \frac{1}{s^2}\Big)^{-\frac{3}{2}}\Big] \\ 
= \frac{\beta^2 (\eta\eta')^{q-1}}{8\pi^2}\Big[finite\ term + \int_{N}^{\infty}ds\ s^{-\frac{5}{2} -\nu}\Big(1 -\frac{3}{2} \bigg(-2\frac{Z}{s} + \frac{1}{s^2}\bigg) + \frac{3*5}{2*2*2}\bigg(-2\frac{Z}{s} + \frac{1}{s^2}\bigg)^2 + ...  \Big)\Big] \\ 
= \frac{\beta^2 (\eta\eta')^{q-1}}{8\pi^2}\Big[finite\ term + \Big(\frac{s^{-\frac{3}{2}- \nu}}{-\frac{3}{2}- \nu} + 3 Z \frac{s^{-\frac{5}{2}- \nu}}{-\frac{5}{2}- \nu} -\frac{3}{2}\frac{s^{-\frac{7}{2}- \nu}}{-\frac{7}{2}- \nu} + (lower\ powers\ of\ s) \Big) |_{N}^{\infty}\Big]
\end{multline}
Since $Z = 1 + \frac{(\eta - \eta')^2 - (\Delta \vec{x})^2}{2\eta\eta'}$, we see that  the highest collective power of $\eta$ and $\eta'$ is $-3 - 2v$  and one such highest power term is multiplying an $\eta$ and $\eta'$ independent and always diverging term $s^{-\frac{3}{2}- \nu}|_{N}^{\infty}$ in the expression for Wightman function. This implies that the behaviour of noise kernel, in this case, is same as for the $\nu = -\frac{3}{2}$ case (because the divergences are determined by the highest power(for $\eta \to \infty$), and hence most divergent term.).  In fact, we have 
\begin{equation}
\langle t_{00}(\eta,\vec{x})t_{00}(\eta,\vec{x'}) \rangle_{P.L.} = \frac{H^{4q}(q-1)^4}{128 \pi^4 \eta^{8-4q}\epsilon^2} + O(\epsilon^{-1})
\end{equation}
where $q = \nu -\frac{1}{2}$ and $\frac{1}{\epsilon} = \frac{s^{-\frac{3}{2}- \nu}|^{\infty}}{\frac{3}{2} +\nu}$ .
\subsection*{Case $\nu > \frac{3}{2}$:}
Now, consider the integral for small $s$ values i.e., 
\begin{multline}
G(x,x')= \frac{\beta^2 (\eta\eta')^{q-1}}{8\pi^2}\Big[finite\ term + \int_{0}^{\epsilon}ds\ s^{\frac{1}{2} -\nu}\Big(1 - 2 Z s + s^2\Big)^{-\frac{3}{2}}\Big] \\ 
= \frac{\beta^2 (\eta\eta')^{q-1}}{8\pi^2}\Big[finite\ term + \int_{0}^{\epsilon}ds\ s^{\frac{1}{2} -\nu}\Big(1 -\frac{3}{2} \bigg(-2 Z s + s^2\bigg) + \frac{3*5}{2*2*2}\bigg(-2 Z s  + s^2 \bigg)^2 + ...  \Big)\Big] \\ 
= \frac{\beta^2 (\eta\eta')^{q-1}}{8\pi^2}\Big[finite\ term +\Big(\frac{s^{\frac{3}{2}- \nu}}{\frac{3}{2}- \nu} + 3 Z \frac{s^{\frac{5}{2}- \nu}}{\frac{5}{2}- \nu} -\frac{3}{2}\frac{s^{\frac{7}{2}- \nu}}{\frac{7}{2}- \nu} + (higher\ powers\ of\ s) \Big) |_{0}^{\epsilon}\Big]
\end{multline}

\bibliographystyle{./utphys1}

\end{document}